\documentclass[twocolumn,showpacs,prb,amsmath,amssymb,floatfix]{revtex4}
\usepackage{bm}

\usepackage{graphicx}
\usepackage{amsmath,amssymb}
\usepackage{bm}
\newcommand{\nn}{\nonumber}

\begin{document}

\title{Tunable transmittance in anisotropic 2D materials}

\author{Phusit Nualpijit$^{1,2}$, Andreas Sinner$^2$, and Klaus Ziegler$^2$}
\affiliation{$^1$Department of Physics, Kasetsart University, Thailand,\\
$^2$Institut f\"ur Physik, Universit\"at Augsburg, Germany}

\begin{abstract}

A uniaxial strain applied to graphene-like materials moves the Dirac nodes along the boundary of the Brillouin zone. An extreme case is the merging
of the Dirac node positions to a single degenerate spectral node which gives rise to a new topological phase. Then isotropic Dirac nodes are replaced by a node with a linear behavior in one and a parabolic behavior in the other direction. This anisotropy influences substantially the optical properties.
We propose a method to determine characteristic spectral and transport properties in black 
phosphorus layers which were recently studied by several groups with angle-resolved 
photoemission spectroscopy, and discuss how the transmittance, 
the reflectance and the optical absorption of this material can be tuned.
In particular, we demonstrate that the transmittance of linearly polarized incident light 
varies from nearly 0\% to almost 100\% in the microwave and far-infrared regime.

\end{abstract}


\maketitle

\section{Introduction}

Since the discovery of graphene~[\onlinecite{novoselov05}], transport properties of different semimetallic materials which crystallize 
on a hexagonal lattice and have Dirac-like electronic spectra, have been in the focus of intensive research. Because of the 
subtle issue of Klein tunneling, these materials turn 
out to have nearly the same conductivity in the broad wave length region ranging from the microwave 
to the visible spectrum~[\onlinecite{nair-sci320,basov08}]. 
An intriguing further development is related to a continuous deformation of the hexagonal lattice,
which can result in a topological transition of the spectrum. For the latter it is crucial that
the six-fold isotropy of the hexagonal lattice is broken by the deformation.
For example, changing the carbon bonds of graphene in one direction (as depicted on the left of Fig.
\ref{fig:CondOm}) changes 
the positions of the two Dirac nodes in the perpendicular direction in momentum space.
This can even lead to their degeneracy in one point. Then the Dirac cones are also affected, since 
the resulting spectrum is linear only in the direction of the bond change
but it becomes parabolic in the perpendicular direction. This case is topologically 
different from a single Dirac node, since the corresponding wave function has a zero winding number, 
contrary to the winding number $\pm 1$ of a single Dirac node. In this sense the system 
undergoes a topological transition, which has a substantial effect on electronic 
and optical properties. These ideas have a long 
history~[\onlinecite{hasegawa06,Dietl2008,montambaux09,Pereira2009,montambaux09a,delplace10,faye14,
Montambaux2016,Sandler2016,Carpentier2017}]. 
More recently, a detailed experimental analysis of black phosphorus layers has revealed that
the spectral properties can be tuned by doping~[\onlinecite{Kim2015,Kim2017,Ehlen2018}]. Using
angle-resolved photoemission spectroscopy (ARPES), it was found that there exists a split
pair of Dirac nodes that can be moved upon doping towards each other along the zigzag direction 
of the underlying honeycomb lattice. The detailed mechanism for the movement of the Dirac nodes in black phosphorus is more complex (cf. Ref.~[\onlinecite{Kim2017a}]) but is based on the fundamental 
mechanism of breaking a discrete lattice isotropy, very similar to the case of an anisotropic
honeycomb lattice. In the following we propose a method
based on light transmittance and absorption which could provide an alternative to ARPES for the observation 
of the moving Dirac nodes in black phosphorus layers. This could serve as a novel analytic method as well as an
application of the moving Dirac nodes for sensors that are sensitive to polarization.  

The intimate connection between electrical 
and optical properties in semimetals allows to study them together. Because of the nearly frequency independent conductivity of 
isotropic graphene its transparency is frequency independent too~ [\onlinecite{nair-sci320,basov08}]. 
For small uniaxial lattice deformations a variation of the transmittance with respect to the wave plane polarization 
has been reported in Refs.~[\onlinecite{{Pereira2010,Oliveira2017}}]. The observed deviation from the isotropic case was small, though,
and did not exceed 1\%. To extend these studies, we will consider stronger lattice deformations in the regime, where the
Dirac nodes are merging. In this case the optical transparency is much more affected and can reach nearly 100\% in the microwave regime. 
It is caused by a divergent optical conductivity in the direction of the stronger bonds and a vanishing optical conductivity
in the direction of the weaker bonds for low frequencies~[\onlinecite{EPL119}]. 
This property opens a possibility for the development of a new type of single atom thick optical polarization filters.

\section{Model}

\begin{figure*}[t]
\includegraphics[width=5cm]{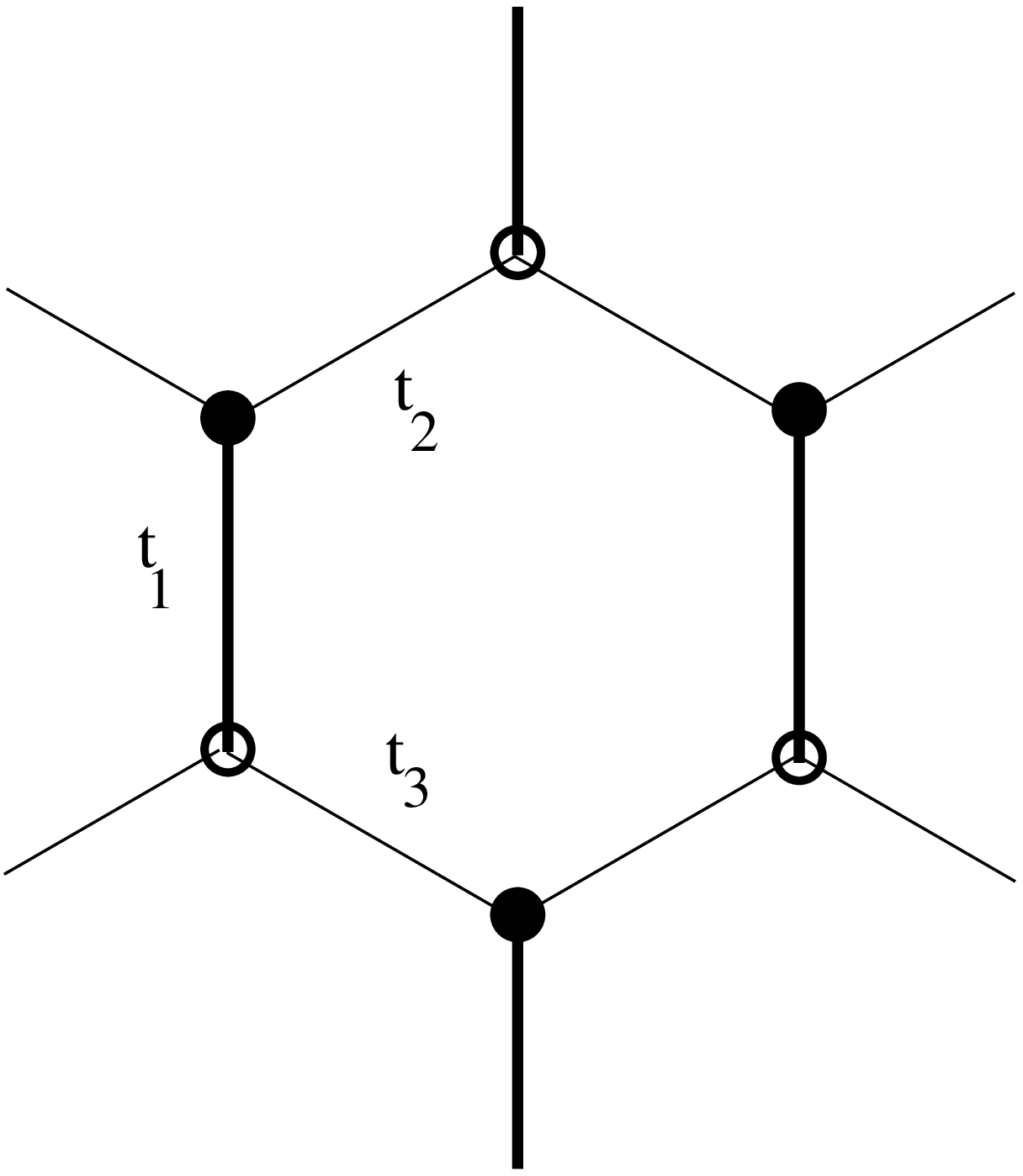}
\hspace{15mm}
\includegraphics[width=8cm]{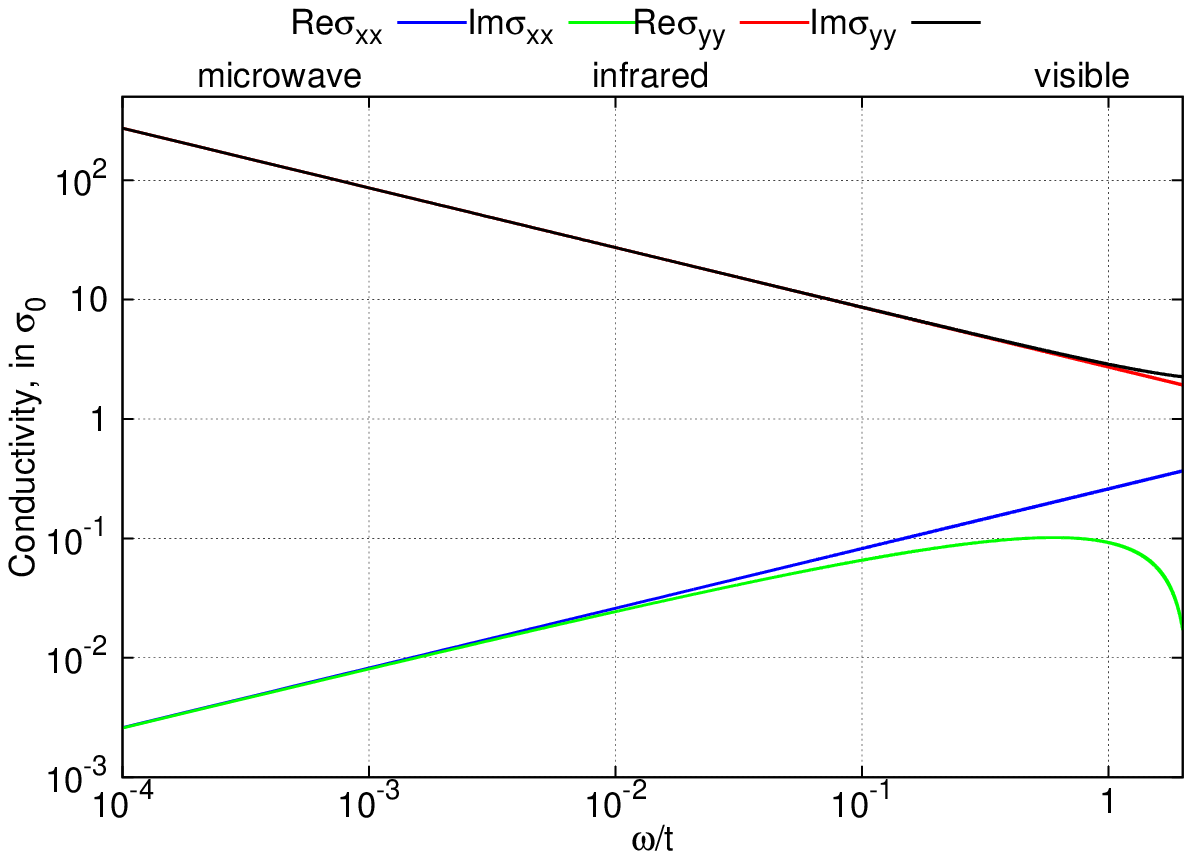}
\caption{Left: Generic hexagonal lattice. The isotropic case is characterized by all equal hopping integrals $t^{}_j$. This isotropy is lifted once $t^{}_j$
become different.
Right: Real and imaginary part of conductivity components. The number $t\sim2.8$eV refers to 
the hopping parameter of the isotropic graphene. The cutoff energy is chosen at the position of 
the van-Hove singularity in isotropic graphene, i.e. $\epsilon^{}_c=2t$.}
\label{fig:CondOm}
\end{figure*}

The tight-binding Hamiltonian for electrons hopping between sites of a hexagonal lattice reads in momentum space
\begin{equation}
\label{eq:TBH}
H = -\sum_{j=1}^{3}
\begin{pmatrix}
0 &  t_j e^{i {a}_j \cdot k } \\
t_j e^{-i {a}_j \cdot k } & 0
\end{pmatrix},
\end{equation}
where the positions of the nearest neighbors around an atom at the origin of coordinates are given by $a_1=a(0,-1)$ 
and $a_{2,3}=a(\pm \sqrt3,1)/2$; $t_j$ denote hopping energies in each direction, cf. Figure~\ref{fig:CondOm}, 
and $a$ represents the distance between nearest neighbors. The spectrum of (\ref{eq:TBH}) has two branches 
corresponding to positive and negative energies
\begin{equation}
E = \pm |t_1 e^{i {a}_1 \cdot {k} } + t_2 e^{i {a}_2 \cdot {k} } + t_3 e^{i {a}_3 \cdot {k} }|.
\end{equation}
For an isotropic lattice, i.e., in the case where all $t^{}_j$ are equal, both spectral branches 
touch each other at six corners of the hexagonal Brillouin zone, where they compose two Dirac nodes. 
An expansion in powers of small momentum deviations around these nodal
points yields in leading order a linear spectrum. The situation changes if the isotropy is lifted, i.e., 
when the hopping integrals are different.
Then the Dirac nodes can be continuously moved in momentum space~[\onlinecite{Dietl2008,montambaux09,Pereira2009,montambaux09a,delplace10,faye14,Montambaux2016,Sandler2016}], 
which also changes the shape of the Brillouin zone, while keeping its area constant.
In the following we will study the case where the hopping integrals $t^{}_2$ and $t^{}_3$ are kept 
fixed at the value of the isotropic lattice $t$, while $t^{}_1$ changes smoothly between $t$ and $2t$~[\onlinecite{EPL119}]. 
Then the nodes (corresponding to Dirac particles with different chirality) 
start moving in the momentum space towards each other: 
\begin{equation}
\label{eq:traj} 
k^{}_x = \pm\frac{2}{\sqrt{3}} \arccos\left[\frac{t_1}{2t}\right],\;\; k^{}_y = \pm\frac{2\pi}{3}
\ .
\end{equation}
At the particular value $t^{}_1 = 2t$ the Dirac nodes merge and give rise to an anisotropic 
spectrum with parabolic dispersion along the direction of the motion and linear perpendicular to it. 
The effective low-energy Hamiltonian then reads
\begin{equation}
\label{eq:EffHam}
H = -\frac{k_x^2}{2m} \hat\sigma_x \pm  c  k_y \hat\sigma_y
\ ,
\end{equation}
where $\hat\sigma_{x,y}$ denote Pauli matrices, $m=2/(3ta^2)$ and $c=3at$. The sign ambiguity in the 
second term is due to different chirality of merging Dirac nodes. The merging point is occupied simultaneously 
by both copies with opposite chiralities. In our subsequent calculation of the 
conductivity, which does not depend on the sign in Eq.~(\ref{eq:EffHam}), we take into account this 
degeneracy by an additional factor of two and assume the sign to be $+$. The effective Hamiltonian has the eigenvalues 
\begin{equation}
\epsilon_{k}= \pm ~\sqrt{\displaystyle \frac{k_x^4}{4m^2}   +  c^2  k_y^2},
\end{equation}
and the corresponding normalized eigenfunctions
\begin{equation}
\psi^{}_{\pm} = \mp \frac{e^{-i k \cdot r}}{\sqrt{2}\epsilon^{}_k}
\begin{bmatrix}
\displaystyle \frac{k_x^2}{2m} + i c k_y \\
\\
\mp\epsilon^{}_k 
\end{bmatrix}.
\end{equation}
The current operators $\displaystyle j_{\mu} = i[H,r_{\mu}]$, corresponding to the 
anisotropic Hamiltonian in Eq.~(\ref{eq:EffHam}), read:
\begin{equation}
\label{eq:Currents} 
j^{}_x = -\frac{k^{}_x}{m}\hat\sigma^{}_x, \;\;\; j^{}_2 = c\hat\sigma^{}_y.
\end{equation}
Then the interband current matrix elements read:
\begin{eqnarray}
\label{eq:MatrElts} 
\langle\psi^{}_\pm|j^{}_x|\psi^{}_\mp\rangle &=& \mp i\frac{c}{m}\frac{k^{}_xk^{}_y}{\epsilon^{}_k},\\
\langle\psi^{}_\pm|j^{}_y|\psi^{}_\mp\rangle &=& \pm i\frac{c}{2m}\frac{k^{2}_x}{\epsilon^{}_k},
\end{eqnarray}
which will be used to compute the conductivity.

\section{Optical conductivity}

To calculate the conductivity per valley and spin projection we use the Kubo formula 
\begin{eqnarray}
\nn
\sigma^{}_{\mu\nu}(\omega) &=& 16i\sigma^{}_0 \sum_{\lambda,\lambda^\prime=\pm}\int\frac{d^2k}{(2\pi)^2} ~
\frac{f_{\beta}(\epsilon^{}_{k,\lambda^\prime})-f_{\beta}(\epsilon^{}_{k,\lambda})}{\epsilon^{}_{k,\lambda}-\epsilon^{}_{k,\lambda^\prime}} \\
\label{eq:kubo}
&&\times~\frac{\langle\psi^{}_\lambda|j^{}_{\mu}|\psi^{}_{\lambda^\prime}\rangle\langle\psi^{}_{\lambda^\prime}|j^{}_\nu|\psi^{}_\lambda\rangle}
{\epsilon^{}_{k,\lambda}-\epsilon^{}_{k,\lambda^\prime} + \omega - i0^+}, 
\end{eqnarray}
Here, $f_{\beta}(\epsilon)=(1 + \exp[\beta(\epsilon-\epsilon_F)])^{-1}$ denotes the Fermi function at inverse temperature $\beta=1/k_BT$ with Boltzmann constant $k_B$. The conductivity in Eq.~(\ref{eq:kubo}) is measured in units of universal dc conductivity per valley and spin projection $\sigma_{0} = 1/16 e^2/\hbar$. Because of the preserved time-reversal symmetry of the model, the non-diagonal elements of the conductivity tensor with $\mu\neq\nu$ are zero. In the zero-temperature limit the diagonal elements of the conductivity tensor read 
\begin{equation}
\label{eq:kubo_t0}
\sigma^{}_{\mu\mu}(\omega) = i\int\frac{d^2k}{(2\pi)^2} \left[ \frac{\Gamma^{}_{\mu\mu}}{2\epsilon_k + \omega+i0^+}   -\frac{\Gamma^{}_{\mu\mu}}{2\epsilon_k - \omega - i0^+}\right].
\end{equation}
Because of the anisotropy, the conductivities in $x$ and $y$ directions are different.
We obtain the following matrix elements of the current-current correlator:
\begin{eqnarray}
\Gamma^{}_{xx} &=& 8\sigma^{}_0 \left(\frac{c}{m}\right)^2\frac{k_x^2 k_y^2}{\epsilon_k^3} ,\\
\Gamma^{}_{yy} &=& 8\sigma^{}_0 \left(\frac{c}{2m}\right)^2 \frac{k_x^4}{\epsilon_k^3}.
\end{eqnarray}
Rewriting momentum integrals as $\int_{-\infty}^{\infty} dk^{}_i \to 2 \int_{0}^{\infty} dk^{}_i$ (which is possible since the integrand function is even under mirroring $k^{}_i\to-k^{}_i$), introducing new integration variables $u=k^2_x/2m$ and $v=ck^{}_y$, and changing into polar coordinates we arrive at
\begin{eqnarray} 
\nn
\sigma_{xx}(\omega) &=& \frac{i\gamma\sigma^{}_0}{(2\pi)^2} \int_0^{\pi/2}d\phi~\cos^{\frac{1}{2}}\phi~\sin^2\phi \\
&\times&\int^{x^{}_c}_0 dx~\sqrt{x}\left[\frac{1}{x - 1 + i0^+} - \frac{1}{x + 1 - i0^+}\right], 
\hspace{0.5cm}\\
\nn
\sigma^{}_{yy}(\omega) &=& \frac{i\gamma\sigma^{}_0}{(2\pi)^2} \int_0^{\pi/2}d\phi~\cos^{\frac{3}{2}}\phi\\
&\times&\int^{x^{}_c}_0\frac{dx}{\sqrt{x}}~\left[\frac{1}{x-1+i0^+}  + \frac{1}{x+1-i0^+}\right],
\hspace{0.5cm}
\end{eqnarray}
where $\gamma=\sqrt{\omega/mc^2}$ and $x=\sqrt{2\epsilon/\omega}$. Angular parts represent standard elliptic integrals and can be found in the literature:
\begin{equation}
 \int_0^{\pi/2}d\phi~{\cos^{\frac{1}{2}}\phi}~\sin^2\phi \sim \frac{1}{2},\;\; \int_0^{\pi/2}d\phi~\cos^{\frac{3}{2}}\phi \sim \frac{7}{8}. 
\end{equation}
Separating real and imaginary parts of the fractions under the integral using the Dirac identity
\begin{equation}
\frac{1}{x \pm 1 \pm i 0^+} = {\cal P}\frac{1}{x \pm 1} \mp i\pi\delta(x \pm 1), 
\end{equation}
where $\cal P$ denotes the operator of the principal part integration. We ultimately obtain 
\begin{equation}
\label{eq:ReSigma}
{\rm Re}~\sigma^{}_{xx}(\omega)  \sim \frac{2\gamma\sigma^{}_{0}}{\pi}   , \;\;
{\rm Re}~\sigma_{yy}(\omega) \sim  \frac{7\sigma^{}_{0}}{2 \pi \gamma}   ,  
\end{equation}
and 
\begin{eqnarray}
\label{eq:ImSigmaXX}
{\rm Im}~\sigma^{}_{xx}(\omega) &\sim& \frac{2\gamma\sigma_{0}}{\pi^2}\left[\ln\left|\frac{x^{}_c-1}{x^{}_c+1}\right| + 2\arctan{x^{}_c} \right],
\\
\label{eq:ImSigmaYY}
{\rm Im}~\sigma^{}_{yy}(\omega) &\sim& \frac{7\sigma_{0}}{2\pi^2\gamma}\left[\ln\left|\frac{x^{}_c-1}{x^{}_c+1}\right| - 2\arctan{x^{}_c}\right],
\end{eqnarray}
as the real and the imaginary part of the conductivity. 
$x^{}_c=\sqrt{2\epsilon^{}_{c}/\omega}$ denotes the effective dimensionless band width. 
At energies $\omega\ll\epsilon^{}_{c}$ we can approximate the expression in the square brackets by 
$\pi/2$ and obtain the same conductivity amplitudes for both real and imaginary part in each direction.
Remarkably, in the low energy regime both conductivities turn out to be functions of the variable
$\gamma=\sqrt{\omega/mc^2}$, taken to mutually inverse powers, though. Hence, the product of both conductivities
does not depend on the frequency, nor on the parameters of the anisotropic Hamiltonian of
Eq.~(\ref{eq:EffHam}). It is related to the universal conductivity per spin projection 
$\bar\sigma^{}_0 = 2\sigma^{}_0$, Ref.~[\onlinecite{EPL119}] 
\begin{equation}
\label{eq:const}
2\lim_{\omega\to 0}\sqrt{|\sigma_{xx}(\omega)| ~|\sigma^{}_{yy}(\omega)|} \sim 1.2 \bar\sigma_0, 
\end{equation}
where the valley degeneracy is taken into account by the factor $2$. The existence of such a relation is dictated by the 
current conservation condition which must hold for any smooth lattice deformations which respects parity, time reversal 
and charge conjugation symmetries. The deviation from unity is due to the low-energy approximation, 
while numerical evaluations with the full tight-binding spectrum gives a value much closer to unity. 

\begin{figure*}[t]
\includegraphics[width=8cm]{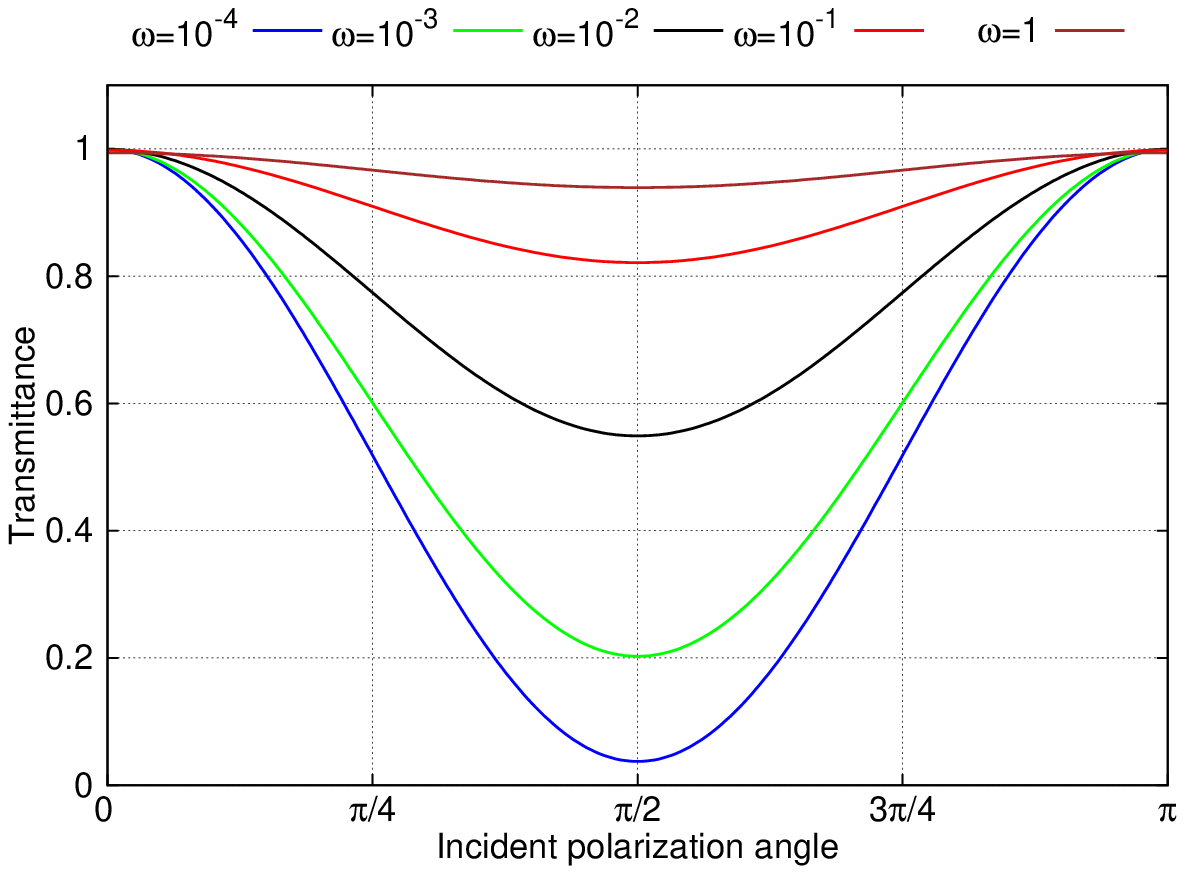}
\hspace{15mm}
\includegraphics[width=8cm]{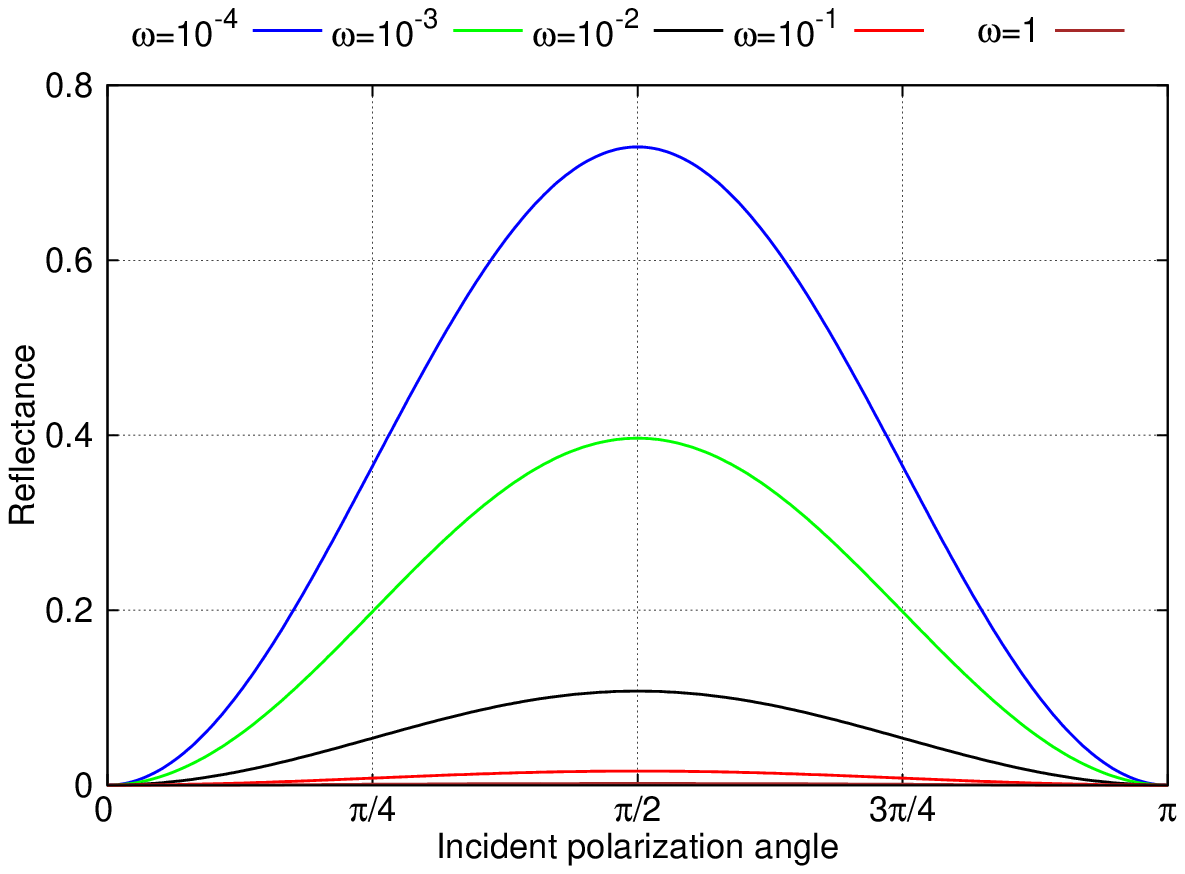}
\caption{Left: Transmittance as function of the incident polarization angle for different light frequencies measured in units of the hopping energy $t$.
Right: Reflectance as function of the incident polarization angle for different light frequencies measured in units of the hopping energy $t$.}
\label{fig:ReInc}
\end{figure*}

\section{Optical properties of strained graphene-like materials}

To study optical properties we must consider the coupling of the electrons in the hexagonal lattice and an external monochromatic
electromagnetic field with frequency $\omega$. 
For this case we have to solve the classical Maxwell equations for the electromagnetic field together
with Ohm's law due to conductivity obtained above. We consider linearly polarized light propagating 
perpendicular to the layer along $z$-direction and the interaction with electrons in the graphene-like 
material at $z=0$. $\phi$ is the angle between the polarization plane and  $x$-axis.  
Then the Maxwell equation of the electromagnetic field reads~[\onlinecite{Quinn1976}]
\begin{equation}
\label{eq:Maxwell}
\frac{\partial^2 \textbf{E}}{\partial z^2}=  \epsilon \frac{\omega^2}{c^2_0}\textbf{E}-4\pi i 
\frac{\omega}{c^2_0} \delta(z) ~\textbf{j}
\end{equation}
with the electronic current in the lattice and $c^{}_0$ denoting the speed of light in vacuum. 
A unique solution of 
this equation is obtained if we assume that the tangential component of electric field is continuous
at $z=0$:
\begin{eqnarray}
\label{sum1}
\tilde{\textbf{E}}^{}_i+\tilde{\textbf{E}}^{}_r = \tilde{\textbf{E}}^{}_t,
\end{eqnarray}
with $i,r$ and $t$ denoting {\it incident}, {\it reflected} and {\it transmitted}, respectively, and $\tilde{\textbf E}$ denoting the amplitude of the corresponding field component, 
along with a discontinuity of its spatial derivative, cf. Appendix 
\begin{equation}
\label{eq:discont}
-iq \tilde{\textbf{E}}_t + iq (\tilde{\textbf{E}}_i-\tilde{\textbf{E}}_r) =4\pi i \frac{\omega}{c^2_0}  ~ \textbf{j}. 
\end{equation}
Exploiting Ohm's law  $\textbf{j}=\sigma \tilde{\textbf{E}}$ and the dispersion relation $\omega=c^{}_0 q$
for the external electromagnetic field, 
we finally obtain a relationship between the transmitted and the incident electric field:
\begin{equation}
\tilde{E}_t^\mu  = \frac{\displaystyle \tilde{E}_i^\mu}{\displaystyle 1 + \frac{\pi\alpha}{2} f^{}_{\mu\mu}} ,\;\;\; \mu = x,y
\label{tr.field}  
\ ,
\end{equation}
with $\tilde{E}_{i}^x=|\tilde{E}_{i}|\cos\phi$, $\tilde{E}_{i,t}^y=|\tilde{E}_{i,t}|\sin\phi$. Here $\alpha=e^2/\hbar c^{}_0\sim1/137$ is the  fine structure constant, $f^{}_{\mu\mu}=\sigma^{}_{\mu\mu}/\sigma^{}_0$ and the valley and spin degeneracy is taken into account by the factor 4. 
The transmittance is defined as the ratio of the intensities of the transmitted and the incident field
\begin{eqnarray}
\nn
T &\equiv&  \frac{I_t}{I_i}  = \frac{|E_t^{x}|^2 + |E_t^{y}|^2}{|E_i^{x}|^2 + |E_i^{y}|^2} \\
&&=T_{x} \cos^2{\phi}+T_{y} \sin^2{\phi}\ ,
\end{eqnarray}
with 
\begin{equation}
T^{}_{\mu} = \left|1 + \frac{\pi\alpha}{2}f^{}_{\mu\mu}\right|^{-2}. 
\end{equation}
The reflectance reads according to Eqs. (\ref{sum1}) and (\ref{tr.field})
\begin{eqnarray}
\nn
R &\equiv&  \frac{I_r}{I_i}  = \frac{|\tilde{E}_t^{x}-\tilde{E}_i^x|^2 + |\tilde{E}_t^{y}-\tilde{E}_i^y|^2}{|\tilde{E}_i^{x}|^2 + |\tilde{E}_i^{y}|^2} \\
\label{reflectance}
&&= R_{x} \cos^2{\phi}+R_{y} \sin^2{\phi} ,
\end{eqnarray}
with 
\begin{equation}
R^{}_{\mu} = \left(\frac{\pi\alpha}{2}\right)^2|f^{}_{\mu\mu}|^2T_\mu.
\end{equation}
Both, transmittance and reflectance are plotted versus the polarization angle in Figure~\ref{fig:ReInc}.
Then the ratio of the absorbed and incident intensity $A=I_a/I_i$ is 
\begin{eqnarray}
\nn
A &=& 1 - R - T \\ 
\label{eq:abs}
& = & 1-\frac{1+|z_x|^2}{|1+z_x|^2}\cos^2{\phi}-\frac{1+|z_y|^2}{|1+z_y|^2}\sin^2{\phi},
\end{eqnarray}
where $z_\mu=\frac{\pi\alpha}{2}f^{}_{\mu\mu}$, which is plotted in Figure~\ref{fig:AbFreqInc} as function of the polarization angle at fixed 
light frequency and in Figure~\ref{fig:AbFreqInc} as function of the light frequency at fixed 
polarization angle. 

\section{Discussion and conclusions}

In sharp contrast to the case of the isotropic hexagonal 
lattice, lifting the isotropy by a 
smooth lattice deformation changes substantially the transport properties. Essential for
our discussion is that the optical conductivity, which is nearly constant with respect to 
the frequency of the incident field (at least in the low energy regime) in the isotropic 
lattice, becomes strongly frequency dependent. In the case studied in this paper, where the armchair 
oriented hexagonal lattice is compressed along the $y$-axis, the corresponding hopping amplitude $t^{}_1$
increases, as visualized Figure~\ref{fig:CondOm}. This enhances the conductivity parallel to $y$-axis 
and suppresses the transport in the perpendicular direction. In the particular case, where $t^{}_1$ becomes twice the hopping energy 
of isotropic lattice, the dc conductivity in $x$-direction is suppressed and the optical 
conductivity vanishes like $\sim\gamma$. On the other hand, the dc conductivity in 
$y$-direction diverges for small $\omega$ like $\sim 1/\gamma$ (cf. Figure~\ref{fig:CondOm}).
Thus, a finite lattice anisotropy ($t_1=2t_{2,3}$) leads to a very strong transport anisotropy at
small frequencies. This reflects the degeneracy of the Dirac nodes, which is a topological effect
in the spectrum. 

\begin{figure*}[t]
\includegraphics[width=8cm]{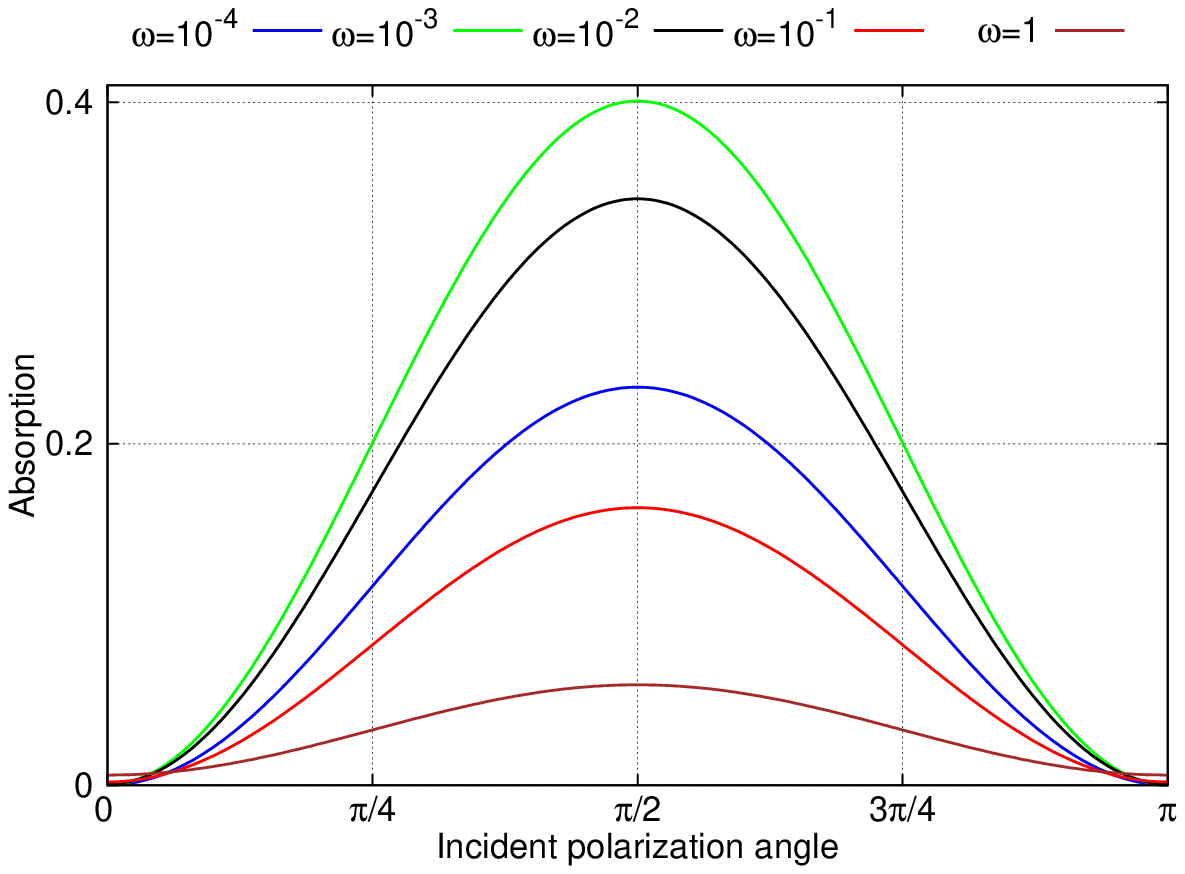}
\hspace{15mm}
\includegraphics[width=8cm]{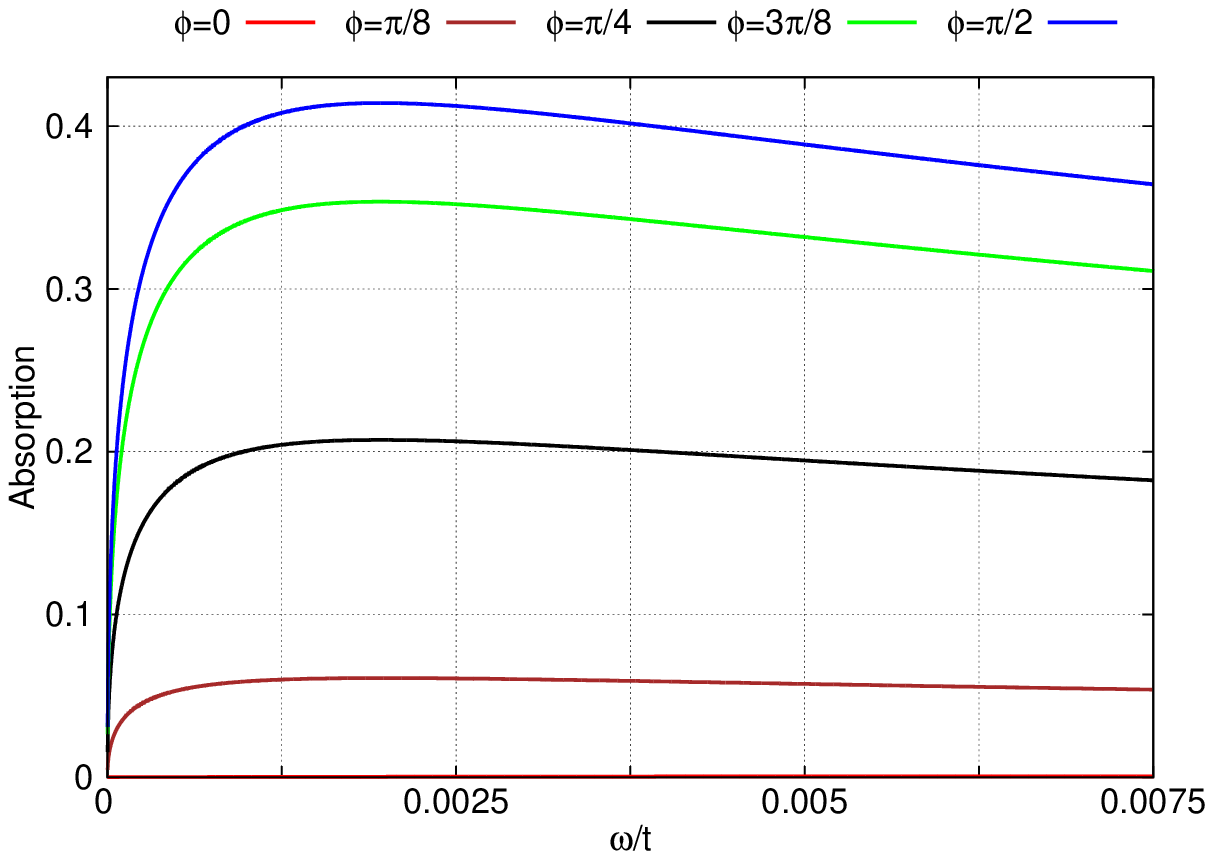}
\caption{Left: Absorption as function of the incident polarization angle for different light 
frequencies measured in units of the hopping energy $t$. Close to $0$ and $\pi$ there are narrow region 
where the absorption is approximately constant.
Right: Absorption as function of the light frequency for different polarization angles.}
\label{fig:AbFreqInc}
\end{figure*}

The strong transport anisotropy has remarkable consequences for the optical properties too, 
in which the transparency and the reflectivity of the system depend on the orientation of the polarization 
of the incident light. For instance, if the polarization is parallel to the $x$-axis, 
the system becomes almost completely transparent in the infrared regime, Figure~\ref{fig:ReInc}. 
This behavior is typical for insulators, which is supported by the fact that $\sigma^{}_{xx}\sim 0$ in the infrared. 
On the other hand, $\sigma^{}_{yy}$ grows towards smaller frequencies and makes the system 
increasingly more reflective for a polarization parallel to the $y$-axis, cf. Figure~\ref{fig:ReInc}. 
Such a behavior is known for conventional metals.  Our calculations indicate that much larger transparency oscillations can occur than those reported 
in Refs.~[\onlinecite{Pereira2010,Oliveira2017}] due to strong deformations with degenerate Dirac nodes.


Another experimentally accessible quantity is the absorption, which is related to the electronic currents induced in the sample. 
This observable quantity has a non-monotonous behavior as a function of both polarization angle and frequency, 
cf. Figure~\ref{fig:AbFreqInc}. In particular, the absorption has a maximum for all frequencies if 
the polarization is parallel to the $y$-axis. However, in the dc limit the absorption goes to zero for any angle $\phi$, 
which is in sharp contrast to the isotropic case, where it is given 
by the constant $\pi\alpha$. This limit is visible in Figure~\ref{fig:AbFreqInc}. With growing frequency,
the absorption initially exhibits a steep growth and reaches a maximum at some specific frequency value.

One could employ the optical method as an alternative to the ARPES studies~[\onlinecite{Kim2015,Kim2017,Ehlen2018}] 
of the spectral anisotropy in black phosphorus layers. This would provide a simple and flexible  approach
either by measuring the surface reflectance, as described in Eq. (\ref{reflectance}) and visualized in Figure~\ref{fig:ReInc},
or by measuring the absorption as presented in Figure~\ref{fig:AbFreqInc}. This method could shed some additional light 
on the peculiar spectral properties found by ARPES. Moreover, it provides not only information 
regarding the spectral properties but also information in terms of transport.

\section*{Acknowledgments}
P.N. acknowledges the financial support by a DPST scholarship of The Institute for the Promotion of 
Teaching, Science, and Technology (IPST), Thailand. A.S. and K.Z. were supported by a grant of the Julian 
Schwinger Foundation for Physical Research.

\appendix

\section{Derivation of the discontinuity condition Eq.~(\ref{eq:discont})}
\label{app:details}

Here we obtain the discontinuity condition Eq.~(\ref{eq:discont}) from the wave equation Eq.~(\ref{eq:Maxwell}). It is obtained with the plane wave ansatz 
\begin{equation}
\label{eq:PlaneWave}
\textbf{E}^{}_{i,t} = \tilde{\textbf E}^{}_{i,t}e^{iqz},\;\;\;\textbf{E}^{}_{r} = \tilde{\textbf E}^{}_{r}e^{-iqz}.
\end{equation}
For this we integrate Eq.~(\ref{eq:Maxwell}) along $z$--axis within an infinitesimally thin region around $z=0$
\begin{equation}
\lim_{\lambda\to 0} \int^{+\lambda}_{-\lambda}dz\left\{\frac{\partial^2\textbf{E}}{\partial z^2} - \epsilon\frac{\omega^2}{c^2_0}\textbf{E} + 4\pi i\frac{\omega}{c^2_0}\delta(z)\textbf{j} \right\} = 0.
\end{equation}
Separating the integral in $z<0$ and $z>0$ parts and noticing 
\begin{equation}
\textbf{E}(z\leqslant0) = \textbf{E}^{}_i + \textbf{E}^{}_{t}, \;\;\; \textbf{E}(z\geqslant0) = \textbf{E}^{}_t,
\end{equation}
and Eq.~(\ref{sum1}) precisely at $z=0$ we get
\begin{eqnarray}
\nn
\lim_{\lambda\to0} \left\{ 
\epsilon\frac{\omega^2}{c^2_0}\int^0_{-\lambda} dz~\left({\textbf{E}}^{}_i+{\textbf{E}}^{}_r\right)
-\epsilon\frac{\omega^2}{c^2_0}\int_0^\lambda dz~{\textbf{E}}^{}_t 
\right.
\hspace{2mm}
\\
\left.+
\left.\frac{\partial}{\partial z}{\textbf{E}}^{}_t\right|^{}_{z=\lambda} -
\left.\frac{\partial}{\partial z}\left({\textbf{E}}^{}_i+{\textbf{E}}^{}_r\right)\right|^{}_{z=\lambda}
\right\} 
= -4\pi i\frac{\omega}{c^2_0}\textbf{j}(0).
\hspace{2mm}
\end{eqnarray}
With the plane wave ansatz Eq.~(\ref{eq:PlaneWave}), terms in the first line of this equation disappear individually in the limit $\lambda\to0$. 
Remaining terms form the condition Eq.~(\ref{eq:discont}).


\begin{thebibliography}{99}
\bibitem{novoselov05}
K.S. Novoselov, A.K. Geim, S.V. Morozov, D. Jiang, M.I. Katsnelson, I.V. Grigorieva, S.V. Dubonos, A.A. Firsov, 
Nature {\bf 438}, 197 (2005).
%
\bibitem{nair-sci320}
R.R. Nair, P. Blake, A.N. Grigorenko, K.S. Novoselov, T.J.
Booth, T. Stauber, N.M.R. Peres, and A.K. Geim, Science {\bf 320}, 1308 (2008). 
%
\bibitem{basov08}
Z.Q. Li, E.A. Henriksen, Z. Jiang, Z. Hao, M.C. Martin, P. Kim, H.L. Stormer, and D.N. Basov, 
Nature Physics {\bf 4}, 532 (2008).
%
\bibitem{hasegawa06}
Y. Hasegawa, R. Konno, H. Nakano, and M. Kohmoto, Phys. Rev. B {\bf 74}, 033413 (2006).
%
\bibitem{Dietl2008}
P. Dietl, F. Pi\'{e}chon, and G. Montambaux, Phys. Rev. Lett.  {\bf 100}, 236405 (2008).
%
\bibitem{montambaux09} 
G. Montambaux, F. Pi\'echon, J.-N. Fuchs, and M.O. Goerbig, Phys. Rev. B {\bf 80}, 153412 (2009).
%
\bibitem{Pereira2009} 
V. M. Pereira, A. H. Castro Neto, and N. M. R. Peres, Phys. Rev. B {\bf 80}, 045401 (2009).
%
\bibitem{montambaux09a} 
G. Montambaux, F. Pi\'echon, J.-N. Fuchs, and M.O. Goerbig, Eur. Phys. J B {\bf 72}, 509 (2009).
%
\bibitem{delplace10} 
P. Delplace and G. Montambaux, Phys. Rev. B {\bf 82}, 035438 (2010).
%
\bibitem{faye14} 
J.P.L. Faye, D. S\'en\'echal, S.R. Hassan, Phys. Rev. B {\bf 89}, 115130 (2014).
%
\bibitem{Montambaux2016} 
P. Adroguer, D. Carpentier, G. Montambaux, and E. Orignac, Phys. Rev. B 93, 125113 (2016).
%
\bibitem{Sandler2016} 
R. Carrillo-Bastos, C. Le{\'{o}}n, D. Faria, A. Latg{\'{e} }, E. Y. Andrei, and N. Sandler, Phys. Rev. B {\bf 94}, 125422 (2016).
%
\bibitem{Carpentier2017} D. Carpentier,  in {\it Dirac Matter}, Progr. Math. Phys., {\bf 71},  Eds.: B. Duplantier, V. Rivasseau, and J.-N. Fuchs, Birkh\"auser (2017).

%
%
%
\bibitem{Kim2015}
J. Kim, S. S. Baik, S. H. Ryu, Y. Sohn, S. Park, B.-G. Park,
J. Denlinger, Y. Yi, H. J. Choi, and K. S. Kim, Science {\bf 349}, 723 (2015).

\bibitem{Kim2017} 
J. Kim, S.S. Baik, S. W. Jung, Y. Sohn, S.H. Ryu, H. J. Choi, B.J. Yang, and K.S. Kim, Phys. Rev. Lett. {\bf 119}, 226801 (2017).
%
\bibitem{Ehlen2018}
N. Ehlen, A. Sanna, B. V. Senkovskiy, L. Petaccia, A.V. Fedorov, G. Profeta, and A. Gr\"uneis,
Phys. Rev. B {\bf 97}, 045143 (2018).

\bibitem{Kim2017a} 
Supplementary Material for
J. Kim, S.S. Baik, S. W. Jung, Y. Sohn, S.H. Ryu, H. J. Choi, B.J. Yang, and K.S. Kim, Phys. 
Rev. Lett. {\bf 119}, 226801 (2017).

%
\bibitem{Pereira2010} 
V.M. Pereira, R.M. Ribeiro, N.M.R. Peres and A.H. Castro Neto, EPL {\bf 92}, 67001 (2010).
%
\bibitem{Oliveira2017}  
O. Oliveira, A. J. Chaves, W. de Paula, and T. Frederico, EPL {\bf 117}, 27003 (2017).
%
\bibitem{EPL119} 
K. Ziegler and A. Sinner, EPL {\bf 119}, 27001 (2017).
%
\bibitem{Quinn1976} 
K.W. Chiu, T.K. Lee, and J.J. Quinn, Surface Science {\bf 58}, 182 (1976).
%
\bibitem{Hill2011} 
A. Hill, A. Sinner and K. Ziegler, New J. Phys. {\bf 13}, 035023 (2011).
%
\end{thebibliography}
\end{document}